\documentclass[helvet,twocolumn]{NaudLab-Class}
\usepackage{blindtext}

\usepackage{mdframed}
\definecolor{shadecolor}{RGB}{220,220,220}

\usepackage{abstract}

\mdfdefinestyle{mdfabstract}{%
    linecolor=blue, linewidth=0.8pt,
    backgroundcolor=shadecolor,
    leftline=false, rightline=false,
    topline=false,
    bottomline=false,
    innertopmargin=0.25cm, innerbottommargin=0.25cm,
    innerleftmargin=0.25cm, innerrightmargin=0.25cm
}

\leadauthor{Naud}

\begin{document}

\title{Connecting levels of analysis in the computational era}
\shorttitle{Connecting levels of analysis}
\author[1,2,3,4,\Letter]{Richard Naud}
\affil[1]{Department of Cellular and Molecular Medicine, University of Ottawa, K1H 8M5, Ottawa, Canada}
\affil[2]{Department of Physics, University of Ottawa, K1H 8M5, Ottawa, Canada}
\affil[3]{Center for Neural Dynamics, University of Ottawa, K1H 8M5, Ottawa, Canada}
\affil[4]{Brain and Mind Institute, University of Ottawa, K1H 8M5, Ottawa, Canada}
\author[1,2,3,4]{Andr\'e Longtin}

\maketitle
\begin{mdframed}[style=mdfabstract]
        \begin{abstract}
            \noindent 
Neuroscience and artificial intelligence are closely intertwined with one another but also with the physics of dynamical systems, philosophy and psychology. Each of these fields tries in its own way to relate observations at the level of molecules, synapses, neurons or behavior to a function. An influential conceptual approach to this end was popularized by David Marr, focused on the interaction between three theoretical 'levels of analysis'. With the convergence of simulation-based methods, algorithm-oriented Neuro-AI and high-throughput data, much of the current research appears organized rather around four levels of analysis: observations, models, algorithms and functions. Bidirectional interactions between these levels influence our undertaking of this interdisciplinary science.
        \end{abstract}
\end{mdframed}

\begin{corrauthor}
\texttt{rnaud\at uottawa.ca}
\end{corrauthor}

\section*{Introduction}

Even if scientific discovery rarely unfolds according to a predictable pattern, the research process may proceed with a recognizable methodology. Every scientist has acquired some intuitions for finding new, potent and relevant knowledge, but these intuitions are often circumscribed by the boundaries of a specific field. What if, as is increasingly recognized in brain science, discoveries rely on inter-disciplinary research? What then are the intuitions? 

\begin{figure*}[!ht]
\centering
\includegraphics[width=1\textwidth]{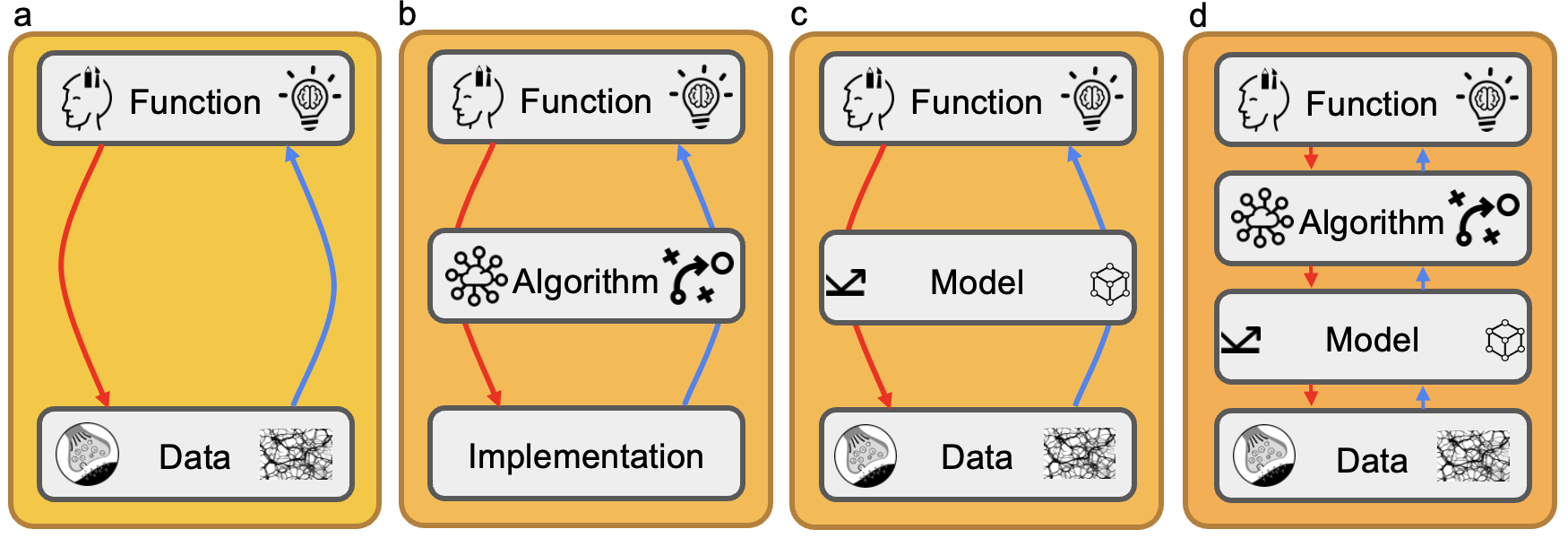}
\caption{Linking levels of analysis through different steps. \textbf{a}. Direct links can be drawn from observations to the associated function (bottom-up, red arrow) or from function to observations (top-down, blue arrow). \textbf{b}. David Marr's \cite{Marr1982a} three levels of analysis link the physical implementation to function, i.e. to the 'computational level of analysis' via algorithms. The function level thus refers to what is being computed. \textbf{c}. Modeling approaches link observations to function by numerical simulations of a mathematical model. \textbf{d}. A linear arrangement of the four levels of analysis. }
\label{fig-LoA}
\end{figure*}

In neuroscience, the quest to relate experimental observations about brain areas, neurons or synapses with underlying functions has been pursued with a number of different approaches (Fig. \ref{fig-LoA}). While a direct relationship between molecular-, cellular- or network-level observations can sometimes be drawn with a particular function, intermediate steps requiring different kinds of expertise are being increasingly recognized as an integral part of our understanding. David Marr and Tomaso Poggio have famously considered three realms -- or levels of analysis -- for the theoretical treatment of a complex neural system. \cite{Reichardt1976a,Marr1982a}. They found useful to separate research on function, termed the computational level, from research on how this function can be achieved, i.e the algorithmic level, from research on how an algorithm can be implemented, i.e. the biophysical implementation  level (Fig. \ref{fig-LoA}b). Interactions between these three separate versions of the same problem have remained a \emph{modus operandi} for some forty years after it was disseminated \cite{Peebles2015a,Eliasmith2015a,Hauser2016a}. Recent developments in machine learning has rejuvenated this approach with the field of Neuro-AI \cite{Richards2019b,Hassabis2017a}, empowering the middle level of analysis with many new algorithms. Data was excluded from these levels but are generally thought to influence every level more or less directly.

Meanwhile, another seemingly different paradigm has long animated efforts in the parts of neuroscience concerned with a more detailed view of the brain and its numerical simulation \cite{Gerstner2012a,Einevoll2013a}. This paradigm aims to link function and observations via an intermediate step consisting of mathematical models of physical entities, e.g. synapses, neurons, neural populations, and brain areas (Fig. \ref{fig-LoA}c). The models at this level of analysis are simulated on computers and/or analyzed with theory. This simulation-driven approach can thus also be seen as involving three levels of analysis: function, models, and data. This is an example of the intense focus on data in computational research, which is justified not only by its provision of constraints to the other levels of analysis, but also by their offer of novel theoretical perspectives arising from their high-throughput or high-dimensional nature. Together, these ideas lead to a four-level approach for relating observations to function (Fig. \ref{fig-LoA}d).

\subsection*{Marr's three levels of analysis}

The path from observations to function is sometimes obvious. At other times, it is obscure. Such is the case for complex concepts like attention, motor learning, perceptual learning, and in fact for almost any putative function of a brain area. The problem appears thorny whether one is approaching it from the observations-to-function direction (from bottom to top in Fig. \ref{fig-LoA}), or from the function-to-observations direction (from top to bottom in Fig. \ref{fig-LoA}).

Among their multiple contributions to neuroscience, Marr and Poggio described a general approach \cite{Reichardt1976a,Marr1982a} to relate observations and function. This approach is exemplified in simple terms in Marr's contribution to the role of the cerebellum in motor control in the 1960s. He investigated the relationship between the function of the cerebellum and its structure. If one approaches the problem from the macroscopic to the microscopic scale, i.e. in the top-down direction, one quickly realizes that there is a functional ambiguity arising from the tuning of motor command sequences. There are so many ways to control muscles or actuators. Not every way is efficient, and not every way is easy or even possible to implement given the biological constraints. There may be no known algorithms for the function suggested by observations. If this is the case, we must construct one \textit{de novo}. Engineering, with its sub-fields of control theory, signal processing and more recently artificial intelligence, provides a generous bank of algorithms for neuroscience \cite{OLeary2014a,Richards2019b}. Given this plurality of options, we would need to settle on well-defined processing steps for motor control. Generating, comparing and modifying the ways in which a function is performed is, in essence, the algorithmic level of analysis. 

Such an algorithmic description can then be related to an implementation. For the cerebellum, an error-based algorithm for motor tuning can then be related to the property of particular cells in this brain area. The biophysical description of a particular implementation was called the 'implementation' level.



\subsection*{Four levels of analysis}

While the algorithmic level of analysis may have sprung from top-down efforts to link levels of analysis, attempting to link these levels from the bottom up, i.e. starting from observations (data), also requires an intermediate step. How do we connect data about neurons, synapses or firing rates to a function? Or to an algorithm? The grand book -- that is the brain -- "now stands open to our gaze but it cannot be understood unless one first learns to comprehend the language in which it is written" \cite{GalileiAssayer}. Like the physics of stellar objects and gravity, the mystery of neural coding and information processing capabilities is written in the language of mathematics. In an attempt to unravel this mystery, this "computational neuroscience" transcribes observations about e.g. synaptic plasticity, molecular interactions, and correlations between brain area activities into mathematical models. Computational units in these models represent idealized physical structures, an association akin to an implementation level of analysis. The model is meant to reproduce known features of the data, to allow in silico experimentation and prediction, and to offer insight into what is being computed. Connecting the language of mathematics to our observations of live neuronal networks thus helps us discern algorithmic elements and infer function; without this link, we remain in one kind of "dark labyrinth"\cite{GalileiAssayer}. 

Algorithms (Fig. \ref{fig-LoA}b) seeking a bridge to data can then rely on mathematical models of observations (Fig. \ref{fig-LoA}c). The search is fostered by this simultaneous visualization through the lens of mathematics of both sides of the river separating data and function.  To facilitate this task, mathematical analysis at the modeling level can map out the behaviour of the system across  parameters \cite{Brunel2000a, Naud2008a, Izhikevich2007a}, separating modes of operation of a complex system. Analytical methods -- when amenable -- make these links particularly clear cut. In fact, the focused development of analytical tools from mathematical physics and mathematics to achieve more abstract representations of what is going on at all levels is sometimes referred to as ``theoretical neuroscience", tacitly acknowledged in Fig. \ref{fig-LoA}d. The comparison between modeling and algorithmic descriptions suggests constraints inspired from the realm of possible implementations. Thus an ideal algorithm can be approximated by processes with biological constraints, constraints inherited from a particular mathematical model. The resulting paradigm is more conveniently reframed using four levels (Fig. \ref{fig-LoA}d).

\begin{figure}[!ht]
\centering
\includegraphics[width=.5\textwidth]{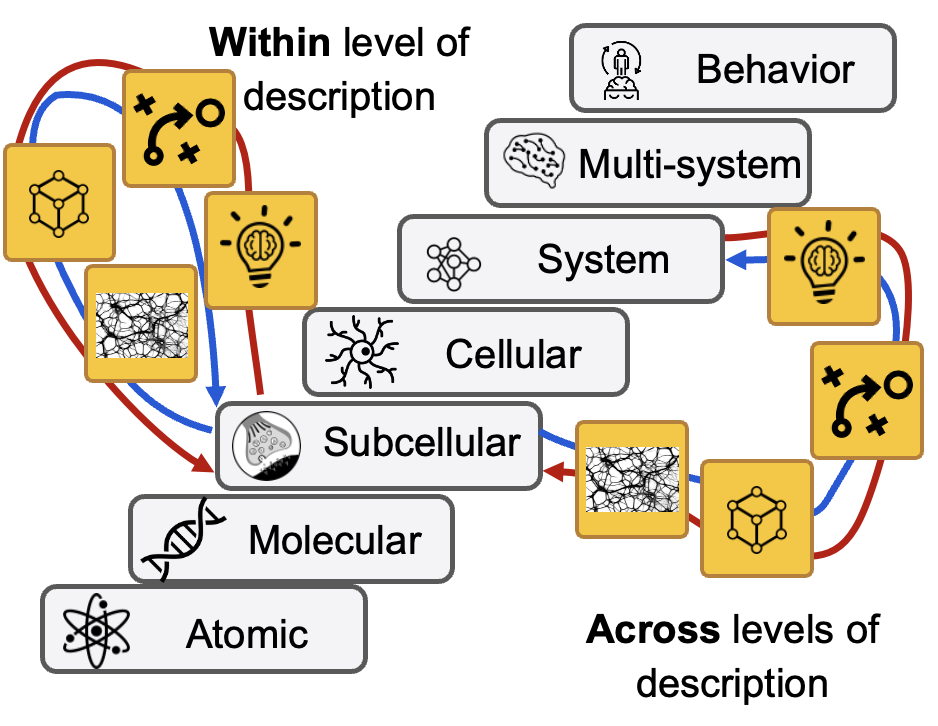}
\caption{Levels of description (gray boxes) can be connected with the help of levels of analysis (yellow squares). }
\label{fig-LoD}
\end{figure}

\subsection*{Data as a level of analysis}

Where do observations enter Marr's scheme? The concept of levels of analysis was meant to link biophysical substrate to the more theoretical algorithmic and functional realms, thereby leaving data unassigned to any level. Today however, much effort is focused on theoretical analysis of data itself, without involving models, algorithms or function. A variety of latent space analyses of raw data can extract meaningful low-dimensional descriptions of high-dimensional data \cite{Yu2008b,Cunningham2014,Pandarinath2018a}.   Other examples include unsupervised clustering of cell types \cite{Hodge2019a} or connection types \cite{Beninger2023a}, and emergent patterns of brain connectivity \cite{Thiebaut2022a}.

Furthermore, formally including the data is essential to see the different levels of analysis as providing constraints to each other \cite{Eliasmith2015a}. For example, observations clearly constrain the implementation level. But it is equally important to consider that the research into functions, algorithms and their implementation influence the type of experiments sought and  performed as well as guide the choice of data analyses.



Observations can sometimes appear to be directly related to an algorithm, without going through a computational model. Algorithms, however, do not contain physically interpretable quantities. Relating physical and  algorithmic quantities relies first on a model of the experimental data. This step is at times so straightforward that it is done tacitly, without thinking that the link arises from a specific model. Therefore, the arrow from "data" in Fig. \ref{fig-LoA}D cannot reach "algorithm" without passing first by "model".

 \section*{Linking levels of analysis at different levels of description}
 
Observations are made at many \emph{levels of description}\footnote{Levels of description are separated according to the spatial and temporal scales of a physical system. They are also know as \emph{levels of granularity}, \emph{hierarchy of scales} or simply \emph{scales}. }, from the molecular to the behavioral or even societal levels, with synapses, cells and brain areas in between (Fig. \ref{fig-LoD}). Levels of description and levels of analysis are different but easily conflated \cite{Mcclamrock1991a}. At each level of description, there are functions and observations.  Both observations and our attempts to understand them can be enclosed within their level of description, a manifestation of dynamical sufficiency \cite{Lewontin1974a} and effective theories  \cite{Burgess2007a}. For example, as much as synapses contribute to behavior, behavior is understood in terms of the interaction of an 'agent', ignoring synaptic dynamics in order to maintain the description within its scale. 

 By attempting to relate observations to function, we may use a number of conceptual frameworks, which form a hierarchy of conceptual scope rather than physical scale. The two hierarchies of interest here, namely, levels of analysis and levels of description, often align. For instance, the function of synaptic plasticity is often framed in terms of selecting network level representations. Explicitly, this is an attempt to link observations (a low level of analysis) of synaptic dynamics (a microscopic level of description) to a function (a high level of analysis) for neuronal networks (a mesoscopic level of description). In some cases, however, the analysis may be circumscribed within a level of description (Fig. \ref{fig-LoD}). For instance, the function of spike-frequency adaptation (a cellular-level observation) is connected to preserving response sensitivity despite a limited dynamic range (a cellular-level function). 

The need to limit modeling to the relevant level of description often leads to phenomenological models that provide a more abstract mathematical description of observations. In turn, research at this level of analysis aims to establish the mathematical regularities within a level of description without addressing the problem associated with linking levels of descriptions. This applies to many features of neural data, ranging from learning rules at the level of populations \cite{Bienenstock1982a}, single cells \cite{Gerstner2018a} and molecules \cite{Maki2020a} to the activity of neuronal populations \cite{Deco2008a} down to that of single-cells \cite{Pillow2008a,Mensi2012a,Pozzorini2013a}. Together, levels of analysis and levels of description are distinct but easily confounded.

\subsection*{Enabling disciplines to work together}

This four-step approach is not the only path that leads from structure to function or the converse. Neither is it necessary to embark on it in order to produce scientific knowledge. We find this organization useful particularly because scientific leaps often depend on multiple disciplines coming together, and the four-step paradigm provides a way to organize collaborations. Observations delve deep into physiology and experimental techniques, which requires experts in these fields. These experts may connect to those diving deep into the modeling of dynamical systems, requiring expertise into a blend of physics, mathematics and high-performance computing. This computational level may then connect with experts at the algorithmic level, requiring expertise in machine learning and engineering. The fourth functional level of analysis may also require expertise in philosophy and psychology, which come in juxtaposition with the mathematics of artificial intelligence when connecting to the level of algorithms. 

Many new graduate and undergraduate programs now train scientists at the intersection between traditional disciplines. But for most, inter-disciplinary science relies on collaborations. In our experience, collaborations are most fruitful when ample time is given to establish an understanding between the distinct but complementary expertise. It takes years for researchers in distinct fields to start understanding the abilities, goals and language of each other. We wrote this comment hoping to provide a motivation for others to boldly set out onto this path.

\end{document}